# ProtoFold Neighborhood Inspector

Nicolas F. Chaves-de-Plaza*, Klaus Hildebrandt* and Anna Vilanova**

*CGV group, TU Delft | **Mathematics and Computer Science, TU Eindhoven

## 1 INTRODUCTION

Post-translational modifications (PTMs) affecting a protein's residues (amino acids) can disturb its function, leading to illness. Whether or not a PTM is pathogenic depends on its type and the status of neighboring residues. In this paper, we present the ProtoFold Neighborhood Inspector (PFNI), a visualization system for analyzing residues neighborhoods. The main contribution is a visualization idiom, the Residue Constellation (RC), for identifying and comparing three-dimensional neighborhoods based on per-residue features and spatial characteristics. The RC leverages two-dimensional representations of the protein's three-dimensional structure to overcome problems like occlusion, easing the analysis of neighborhoods that often have complicated spatial arrangements. Using the PFNI, we explored proteins' structural PTM data, which allowed us to identify patterns in the distribution and quantity of per-neighborhood PTMs that might be related to their pathogenic status. In the following, we define the tasks that guided the development of the PFNI and describe the data sources we derived and used. Then, we introduce the PFNI and illustrate its usage through an example of an analysis workflow. We conclude by reflecting on preliminary findings obtained while using the tool on the provided data and future directions concerning the development of the PFNI.

## 2 DATA

The challenge organizers provided two datasets that describe how PTMs relate to proteins' structures[1]. First, the *structures* dataset includes, for each residue (amino acid) of three proteins across two organisms, the 3D coordinates of its alpha, beta, and carboxyl carbon atoms and its nitrogen atom. It also includes, for each residue, an estimate of the prediction confidence (pLLDT) and the type of structure. Second, the *modifications* dataset lists the PTMs that affect the residues in the *structures* dataset. Each row in the *modifications* dataset describes a PTM that affects a protein´s residue, including the PTM's name/type, classification, and pathogenic status. A residue in the *structures* dataset can be related to multiple modifications.

Finally, we derived a third dataset, *neighbors*, from the *structures* dataset. The *neighbors* dataset lists the thirty nearest neighbors of every residue within every protein structure. We used the Euclidean distance between the residues' alpha carbons as the metric to filter the neighbors.

## 3 MODELING AND TASKS

We define the general goal of the visualization system as enabling the analysis of three-dimensional neighborhoods of proteins' residues. Before proceeding, we formulate this goal explicitly using the language of graph theory.

We model proteins as graphs where residues are nodes indexed by the position in the protein's primary structure. To simplify modeling, we use the position of the residue's alpha carbon as a proxy of the residue's position. In terms of links between nodes, there are two types: topological links between consecutive residues and spatial ones between residues that are close in three-dimensional space. The PFNI focuses on the latter. Finally, nodes and links have attributes. Examples of the former include the number and type of PTMs and the pLLDT score. As for the latter, we consider the Euclidean distance between the nodes that the link connects.

Considering this definition of the problem and the overall goal, we now list the tasks that the PFNI should support:

- (T1) Select nodes (residues) for neighborhood analysis based on node-level and neighborhood-level characteristics.
- (T2) Identify salient patterns in the distribution of a neighborhood's node and link-level characteristics.
- (T3) Compare neighborhoods to assess their similarity, or dissimilarity, under a feature of interest.

## 4 VISUALIZATION SYSTEM: PFNI

Figure 1 presents an overview of the ProtoFold Neighborhood Inspector (PFNI). We developed the prototype using web technologies like D3 [1] and Three.js [2]. The following paragraphs describe the system's main idiom, the Residue Constellation, and the supporting Three-Dimensional and Bulk Selection widgets.

### 4.1 Residue Constellation

#### 4.1.1 Primary Structure Orbits (PSO)

We use a radial layout to organize the protein's two-dimensional primary structure. The main goal of the PSO is to let users identify residues for subsequent neighborhood-level analysis. The PSO consists of three orbits, each consisting of as many arcs of equal length as residues in the protein. Figure 2 presents the three orbits in detail. The inner orbit displays the position of each residue in the chain using a sequential color scale. In the middle orbit, the colors of the arcs change depending on the user-selected categorical variable of interest. Finally, the outer orbit uses the arcs' thickness to encode a user-selected numeric variable. In Figure 2, the numeric variable is the number of modifications each residue has.

Figure 5 shows an additional layer of information that this layer provides. When the user hovers over a residue, ribbons connecting the residue to its neighbors appear. Using this transient ribbon view, the user can get a sense of the residue's neighborhood before adding it to the Neighborhoods Force Layout.

We chose a radial layout to depict the protein's primary structure for two reasons. First, it allows positioning residues in large chains side-by-side, which permits using the orbits' arc's thickness to judge relationships between numerical values. Second, because the center of the PSO remains empty, it allows an efficient overview of the neighborhoods via the ribbon view.

#### 4.1.2 Neighborhoods Force Layout (NFL)

In the center of the Residue Constellation lies the Neighborhoods Force Layout (NFL), which provides a 2D representation of the protein 3D neighborhoods. Explicitly, this view organizes user-selected residues (primary nodes) and their neighbors (secondary nodes) using a force layout, similarly to [3]. The NFL depicts primary and secondary nodes differently to reduce clutter and allow multiple levels of neighborhood analysis, as Figures 3 and 4 illustrate. On the one hand, the NFL uses the neighborhood summarization glyph (following subsection) to depict primary

---

Contact: n.f.chavesdeplaza@tudelft.nl
[1] http://biovis.net/2022/biovisChallenges_vis/

nodes. On the other, it uses smaller circles to plot secondary nodes. Although the NFL could only include primary nodes, adding secondary ones allows us to understand which residues contribute to the summarization glyph and identify shared nodes between neighborhoods.

### 4.1.3 Neighborhood Summarization Glyph (NSG)

The NFL represents user-selected residues using Neighborhood Summarization Glyphs (NSGs). NSGs can summarize residue- and neighborhood-level information, composing several encodings, which are shown at the bottom left of Figure 4. First, the inner circle indicates the pathogenic status of the PTMs in the residue's neighborhood and the prediction confidence (pLLDT). For the former, the fill color indicates if the neighborhood contains (orange) or not (green) a pathogenic modification. As for the latter, the inner circle's stroke thickness encodes the average pLLDT across the residue's neighbors. Second, the NSG divides the area between the inner and outer circles in as many arcs as unique modification types exist among the selected residues. When the user selects more than one residue, the thickness of these arcs encodes the proportion of a given modification type that the residue's neighborhood contains. Finally, the color of the outer circle indicates if the primary node contains (red) or not (green) a pathogenic modification. The same coloring applies to the fill of secondary nodes, which helps spot where pathogenic PTMs lie.

NSGs help users with tasks that require perceiving or discriminating patterns in multivariate data [4]. As a standalone glyph, an NSG permits identifying salient patterns in the distribution of neighborhoods' characteristics (T1). When the user selects multiple residues, the glyphs allow users to assess how similar their neighborhoods are (T2).

## 4.2 Three-Dimensional Widget

Analysts recognize proteins as three-dimensional entities. Therefore, we provide an interactive 3D view of the protein to support the mental model users employ when performing the tasks. The 3D widget (top right panel of Figure 1) depicts the protein's 3D structure using simplified balls and sticks representation [5], which alleviates the occlusion problems of representations like shaded surfaces [6]. Users can pan and rotate the 3D view to understand the protein's spatial extent. Furthermore, when users select residues in the RC or with the bulk selector below, the 3D widget highlights them, providing spatial context.

## 4.3 Bulk Selection Widget

The bulk selection widget (bottom right panel of Figure 1) is a horizontal bar chart that permits simultaneously selecting several residues for neighborhood analysis. The bars' widths encode the number of residues containing that value. The bars' colors match the colors of the middle and outer orbits in the RC and the points in the three-dimensional widget, serving as a legend. Clicking on one of the bars includes all residues that have that value in the current selection, triggering an update of the respective views.

## 5 ANALYSIS WORKFLOW

Having described the components of the PFNI, we now present an example of an analysis workflow using the tool, showing the type of tasks users can perform. For the example, we focus on the human transforming growth factor beta-1 proprotein (P01137), the protein with most PTMs. Below we divide the workflow into residue selection and neighborhood analysis phases.

### 5.1 Residue Selection

The first step when using the PFNI is to select the residues to include in the neighborhood analysis (T1). We start by highlighting those with pathogenic modifications, changing the color of the middle orbit to encode the pathogenic status categorical variable. As Figure 5 shows, we can get a sense of these residues' neighborhoods before selecting them using the ribbon view. Of the five pathogenic residues in P01137, those in positions 44, 109, and 224 have farther-reaching neighborhoods versus those in positions 168 and 222. In terms of the residue types, two are cysteine (44 and 109), two arginine (222 and 224), and one glutamic acid (168). Concerning residues' structure types, there is no discernible pattern with only two sharing types (109 and 168 were of type strand).

### 5.2 Neighborhoods Analysis

The NFL and NSG afford to analyze individual residue neighborhoods and also compare several of them. For inspecting individual residues, Figure 4 shows how changing the extent of the neighborhood permits users to see how the neighborhood composition changes in terms of PTMs. For the residue in the figure, it shows that the pathogenic modification occurs relatively distantly between k=19 and k=29. Also, the main change in this residue's neighborhood in that range of k is the appearance of a pathogenic [5] Carbamyl PTM (brown).

Moving beyond the analysis of individual neighborhoods, Figure 3 shows how the NFL permits NSG-based comparison of several ones. Concretely, it depicts the resulting NFL view after selecting the five residues of P01137 containing a pathogenic modification. In this view, it becomes evident that residues 168, 222, and 224 share a large extent of their neighborhoods. The other two residues, 44 and 109, are independent of the rest and exhibit markedly different PTM distributions. This visualization can help analysts understand if certain PTM distributions are more likely to accompany pathogenic modifications. Finally, the analyst can try different neighborhood sizes, same as with individual residues.

## 6 DISCUSSION AND CONCLUSION

The PFNI can enhance analysts' understanding of PTM data by affording a neighborhood-level perspective. Specifically, the Residue Constellation (RC) idiom is a powerful way to identify, inspect and compare three-dimensional residue neighborhoods. To boost the PFNI's applicability in practical settings [7], we plan on extending it in several ways. First, we plan on allowing more complex filters based on multiple variables or newly derived ones. As for the latter, we could, for instance, calculate the number of neighbors within a given radius or properties of the protein's surface [8]. Second, in the current incarnation of the RC, only visual assessment of glyphs permits comparing neighborhoods. We want to compute pairwise neighborhood similarity metrics like alignment distance or histogram difference [9] and expose them to users' via idioms like heatmaps. Finally, we will add the capability to include multiple proteins in the neighborhood analysis. All proteins share the same residue and PTM alphabets. Combining several of them can enrich the dataset with more pathogenic neighborhoods. Nevertheless, this poses a challenge for the Primary Structure Orbits, which currently can only accommodate a protein at a time.

# FIGURES

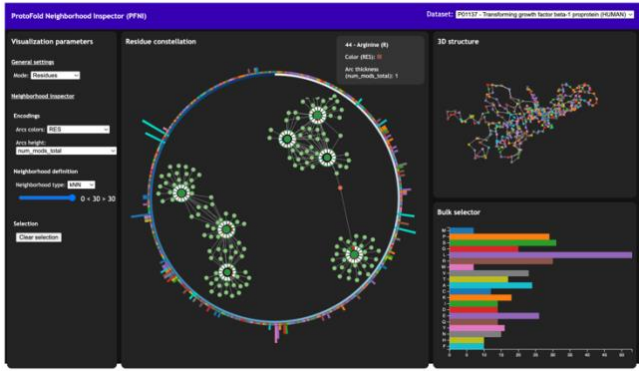

Figure 1: Overview of the ProtoFold Neighborhood Inspector (PFNI). The left panel allows users to modify encodings and define neighborhood selection settings. The right panel includes the principal idiom, the Residue Constellation (RC), and the supporting 3D and bulk selector widgets.

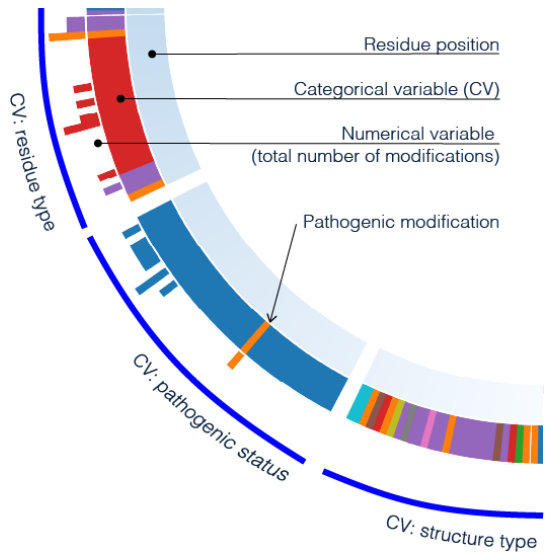

Figure 2: Detail of Primary Structure Orbits (PSO). The PSO permits users to select residues for analysis based on categorical and numerical variables.

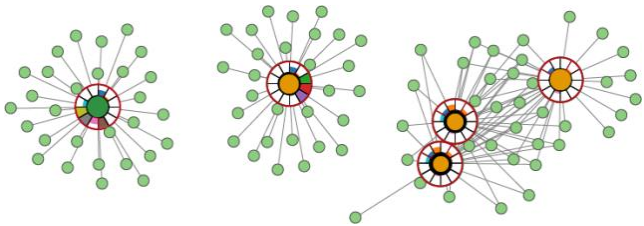

Figure 3: Detail of the Neighborhood Force Layout (NFL). The NFL layout provides a 2D view to inspect the protein's 3D structure via the neighborhood summarization glyphs.

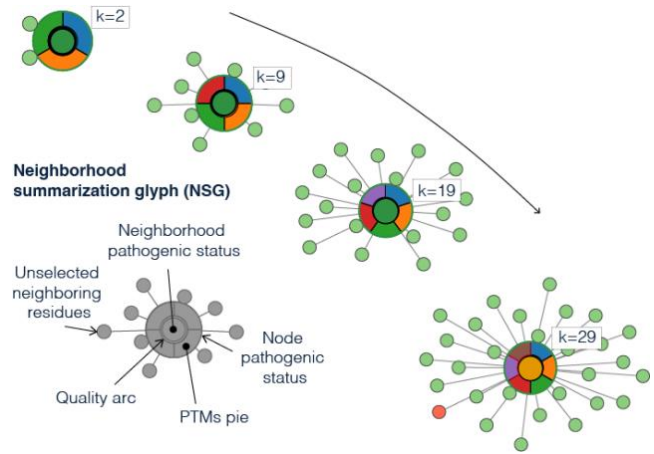

Figure 4: Detail of the Neighborhood Summarization Glyph (NSG). The NSG permits analyzing the distribution of PTMs in a residue's neighborhood. By changing the neighborhoods' extent (k), users can see how the PTM distributions change.

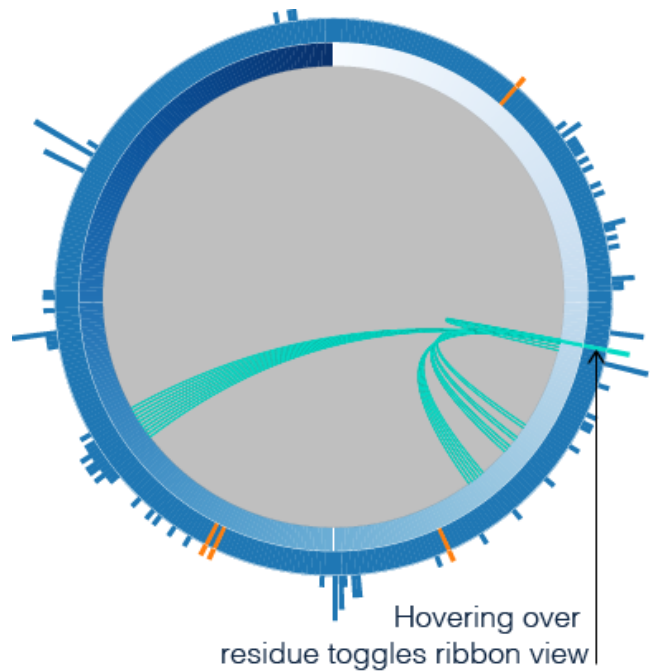

Figure 5: The transient ribbon view provides an overview of a residue's neighborhood in the PSO.